\documentclass[journal]{IEEEtran}
\usepackage{AcronymsIEEE}
\usepackage{glossaries}
\usepackage{Acronyms}

\usepackage{graphicx} 
\usepackage{stfloats}
\usepackage{enumerate}%
\usepackage{graphicx} 
\usepackage{enumerate}%
\usepackage{bm}
\usepackage{multicol}
\usepackage[export]{adjustbox}
\usepackage{lipsum}
\usepackage{mathtools}
\usepackage{cuted}

\usepackage{float}
\usepackage{graphicx} 
\usepackage{caption}
\usepackage[labelformat=simple]{subcaption}

\usepackage[font={footnotesize}]{caption, subcaption}
\usepackage{caption}
\usepackage[labelformat=simple]{subcaption}

\usepackage[font={footnotesize}]{caption, subcaption}
\usepackage{physics}
\PassOptionsToPackage{bookmarks={false}}{hyperref}

\newcommand{\quotes}[1]{``#1''}

\usepackage{AcronymsIEEE}

\PassOptionsToPackage{bookmarks={false}}{hyperref}

\usepackage{amssymb}
\usepackage{bm}

\usepackage[noadjust,compress]{cite}

\usepackage{cite}

\ifCLASSINFOpdf
\else
\fi

\begin{document}

\title{Perturbation-based FEC-assisted Iterative Nonlinearity Compensation for WDM Systems}

%
%

\author{Edson~Porto~da~Silva,~\IEEEmembership{Member,~IEEE},~\IEEEmembership{Member,~OSA}, Metodi P. Yankov, Francesco Da Ros,~\IEEEmembership{Member,~IEEE}, Toshio Morioka,~\IEEEmembership{Member,~IEEE}, Leif K. Oxenl\o we, \IEEEmembership{Member,~OSA}

\thanks{Edson P. da Silva was with the Department of Photonics Engineering (DTU Fotonik), Technical University of Denmark - DTU, \O rsteds Plads 340, Kgs. Lyngby, 2800, now he is with the Department of Electrical Engineering of the Federal University of Campina Grande (UFCG), Paraiba, Brazil. 

Metodi P. Yankov, Francesco Da Ros, Toshio Morioka, and Leif K. Oxenl\o we are with the Department of Photonics Engineering (DTU Fotonik), Technical University of Denmark - DTU, \O rsteds Plads 340, Kgs. Lyngby, 2800, Denmark. Metodi P. Yankov is also with Fingerprint Cards A/S, 2730 Herlev, Denmark. e-mails: edson.silva@dee.ufcg.edu.br, $\lbrace$meya, fdro, tomo, lkox$\rbrace$@fotonik.dtu.dk.}

\thanks{Manuscript submitted 26 of July, 2018; revised .}}

\maketitle

\begin{abstract}
A perturbation-based nonlinear compensation scheme assisted by a feedback from the forward error correction (FEC) decoder is numerically and experimentally investigated. It is shown by numerical simulations and transmission experiments that a feedback from the FEC decoder enables improved compensation performance, allowing the receiver to operate very close to the full data-aided performance bounds. The experimental analysis considers the dispersion uncompensated transmission of a 5$\times$32~GBd WDM system with DP-16QAM and DP-64QAM after 4200~km and 1120~km, respectively. The experimental results show that the proposed scheme outperforms single-channel digital backpropagation.
\end{abstract}


\IEEEpeerreviewmaketitle
\section{Introduction}
\IEEEPARstart{T}HE signal distortions originated from the nonlinear Kerr effects, known as nonlinear interference (NLI), impose limits to the information throughput of wavelength division multiplexing (WDM) systems over single-mode fibers (SMFs) \cite{Ellis2017}. The challenges to overcoming such phenomena have motivated an increasing effort on the investigation of nonlinearity compensation (NLC) techniques. In particular, several digital signal processing (DSP) methods to equalize the nonlinear fiber channel have been proposed \cite{Carledge2017}. Part of the difficulty in dealing with such impairments is due to the large complexity of the signal processing required to equalize the nonlinear fiber channel. 

For an idealized noiseless and deterministic fiber channel, digital backpropagation (DBP) can fully compensate the NLI generated by signal-signal nonlinear interactions happening during propagation, as long as all the frequency components involved are jointly processed \cite{Alic2014, Temprana2015}. Several variants of DBP based on the split-step Fourier method (SSFM) have been studied to compensate signal-signal distortions. 
In practice, the noise originated from the transceivers and the optical amplifiers will also impact the system producing stochastic NLI from signal-noise and noise-noise nonlinear interactions \cite{Ghazisaeidi2017, Galdino2017}. Moreover, signal-signal distortions will also exhibit a degree of randomness due to random fluctuations in the phase or the frequency of the optical carriers \cite{Alic2014}, as well as due to stochastic time-varying effects of the fiber channel, such as polarization mode dispersion (PMD) \cite{Gao2012}. Conventional DBP algorithms do not account for stochastic NLI and their operation resembles a zero-forcing equalization \cite{Proakis2008}. To improve the effectiveness of DBP against stochastic NLI, such algorithms have to be modified \cite{Czegledi2017,Irukulapati2014}. Nevertheless, the stochastic NLI is considered to impose a fundamental limitation to the performance of DBP.

Alternatively to DBP, perturbation-based algorithms can be used to perform NLC. The first-order perturbation analysis of the Manakov equation has been recently investigated as a methodology to design algorithms for intra-channel NLC \cite{Tao2011}. Such algorithms usually operate at one sample per symbol, therefore relaxing sampling requirements when compared to the SSFM. Due to their potential to reduce the DSP complexity, the performance of digital receivers employing such algorithms has been investigated in the literature \cite{Liang2014, Gao2014, Malekiha2016}. 

Perturbation-based NLC algorithms have been mostly employed as transmitter-side pre-distortion techniques since the calculation of the NLI waveform requires the knowledge of the symbols sent through the channel. However, the performance of pre-distortion techniques is bounded by hardware constraints, such as analog bandwidth and the effective number of bits of digital-to-analog converters \cite{Oyama2013, Ghazisaeidi2015}.
Moreover, because the NLI is dependent on the transmitted waveform, pre-distortion is inherently suboptimal. Alternatively, NLC can be realized with a perturbation-based decision feedback equalizer (DFE) at the receiver side \cite{Oyama2014}. However, the efficacy of the post-compensation is bounded by the incomplete knowledge of the receiver on the transmitted symbol sequences. Therefore, at high symbol error rates (SERs), i.e. at low received signal-to-noise ratios (SNRs), the performance of post-compensation can be severely degraded. Hence, as for DBP, the stochastic channel impairments will ultimately limit the performance of the perturbation-based NLC.

The performance of coherent optical receivers is improved by NLC strategies that are adaptive or tailored to track stochastic channel impairments \cite{Yankov2017, Golani2018_1}. Moreover, a performance improvement is expected in receivers that explore the error protection provided by the forward error correction (FEC) codes within the NLC \cite{Sugihara2013}. This potential has been recently explored in the literature. In \cite{Pan2015}, a code-aided scheme has been shown to improve the performance of the expectation-maximization algorithm in mitigating nonlinear phase noise. In \cite{Arlunno2014}, a turbo equalization scheme is proposed for impairment compensation in coherent optical receivers, however only using a normalized least mean square (NLMS) algorithm in the equalization stage.

Intuitively, it is expected that coherent receivers would also benefit from the iteration between FEC decoding and equalization strategies designed according to the physical models of the NLI. In that respect, although perturbation methods are less accurate than SSFM in predicting the NLI distortions, they are better suited for algorithms targeting joint NLC and FEC decoding because they operate at the symbol level.

In this paper, we extend our work in \cite{Silva2018} to investigate the performance of an iterative first-order perturbation-based NLC scheme assisted by feedback from a low-parity density check (LDPC) decoder. Firstly, the proposed NLC scheme is detailed and its performance is numerically investigated via SSFM-based simulations. Secondly, the experimental results presented in \cite{Silva2018} are discussed and extended with an analysis to compare the performance of the proposed scheme with the performance of single-channel DBP for all transmitted WDM channels.

The remaining of the paper is organized as follows. In Section~\ref{SecII}, the perturbation-based NLC methods considered in this paper are described in details and the proposed FEC-assisted iterative scheme is discussed. In Section~\ref{SecIII}, a numerical analysis comparing the performance of different NLC methods with the proposed scheme is shown. In Section~\ref{SecIV}, the analysis presented in Section~\ref{SecIII} is extended to transmission experiments, which is followed by the final remarks.

\section{Perturbation-based Nonlinearity Compensation}\label{SecII}
The perturbation models for the NLI considered in this paper were originally derived for dispersion uncompensated fiber transmission. Therefore, in the following, the analysis and the NLC algorithms presented are restricted to this category of fiber links. Moreover, only single-channel receivers are considered, i.e. the receiver performing NLC detects only one WDM channel.

\subsection{Intra-Channel NLC}\label{IntraNLC}
Consider $\hat{A}_{x}(k)$ to be the detected symbol of polarization-x at the instant $\mathrm{t=kT_s}$, where $\mathrm{T_s}$ is the symbol period. After linear compensation of chromatic dispersion (CD) and matched filtering, $\hat{A}_{x}(k)$ can be expressed as
\begin{equation}\label{Eq1}
\hat{A}_{x}(k) = (A_{x}(k) + \Delta A_{x}(k))\exp(j\phi_{x}(k)) + n_x(k),
\end{equation}
\noindent where $A_{x}(k)$ is the transmitted symbol, $n_x(k)$ is a Gaussian noise process, and $(\Delta A_{x}(k),\phi_{x}(k))$ describe the intra-channel NLI distortion. Here the first-order perturbative approximation of the intra-channel NLI is performed according to the additive-multiplicative model (AM model) described in \cite{Tao2014}.

\begin{figure*}[b!]
\small
\begin{equation}\label{Eq2}
\Delta A_{x} = P_0^{3/2}\left[\sum_{m\neq 0,n\neq 0}[A_x(n)A_x^*(m+n)A_x(m)+A_y(n)A_y^*(m+n)A_x(m)]C(m,n)+\sum_{m\neq 0,n}A_y(n)A_y^*(m+n)A_x(m)C(m,n)\right],
\end{equation}
\begin{equation}\label{Eq3}
\phi_{x} = P_0 \Im\left\lbrace\sum_{m\neq 0} \left( 2|A_x(m)|^2+|A_y(m)|^2 \right)C(m,0) +\left( 2|A_x(0)|^2+|A_y(0)|^2 \right)C(0,0)\right\rbrace, 
\end{equation}
\normalsize
\end{figure*}

The intra-channel NLI waveform parameters are calculated according to Eqs.~(\ref{Eq2})-(\ref{Eq3}), where $\mathrm{P_0}$ is the pulse peak power and $C$ is a matrix of coupling coefficients that depend on the physical parameters of the channel, the pulse shape and baud rate of the transmission \cite{Tao2014}. The double summations in $\mathrm{(m,n)}$ are taken over the symbol intervals $\mathrm{[-L, L]}_x$ and $\mathrm{[-L, L]}_y$. The choice of $\mathrm{L}$ is usually involves a trade-off between how much of the memory present in the channel is incorporated by the model and its computational complexity. Hence, the $\mathrm{(2L+1)\times (2L+1)}$ matrix $C$ corresponds to a discrete model for the intra-channel NLI with a finite memory of $\mathrm{2L+1}$ symbol periods. The same equations apply to the distortions in polarization-y, only exchanging the corresponding indexes.  The indexes in Eqs.~(\ref{Eq2})-(\ref{Eq3}) are relative delays to the symbol at $\mathrm{t=kT_s}$.

In order to use Eqs.(\ref{Eq2})-(\ref{Eq3}) to calculate $(\Delta A_{x/y},\phi_{x/y})$ the receiver has to perform first an estimation on the sequence of transmitted symbols. This operation can be performed via hard decisions (HD) on the received noisy symbols based on the minimum Euclidean distance to a reference constellation. After the estimation of $(\Delta A_{x/y},\phi_{x/y})$, the NLC is performed by subtracting the NLI distortion from the symbol of interest. In this configuration, the perturbation NLC operates similarly to a DFE.

\subsection{Inter-Channel NLC}\label{InterNLC}

When observed from a single-channel receiver, part of the inter-channel NLI can be modeled as a stochastic process that produces time-varying intersymbol-interference (ISI) \cite{Dar2015, Golani2018_2}. Assume $\hat{\bm{A}}(k) = [\hat{A}_{x}(k), \hat{A}_{y}(k)]^T$ to be the detected symbols of both polarizations at $\mathrm{t=kT_s}$. Then, after compensation of CD and intra-channel NLI, $\hat{\bf{A}}(k)$ can be written as
\begin{equation}\label{Eq4}
\hat{\bm{A}}(k) = {\bm{A}}(k) + i\sum_{n} {\bm{H}}_{n}^{(k)}{\bm{A}}(k-n) + {\bm{n}}(k),
\end{equation}
\noindent where ${\bm{A}}(k) = [A_{x}(k), A_{y}(k)]^T$ is the vector of input symbols, ${\bm{H}}_{n}^{(k)}$ is a 2$\times$2 time-varying matrix of ISI coefficients, and ${\bm{n}}(k) = [n_{x}(k), n_{y}(k)]^T$ is a Gaussian noise process. The inter-channel NLI is represented by ${\bm{H}}_{n}^{(k)}$, whose coefficients are functions of the physical parameters of the fiber channel and the data symbols transmitted in the co-propagating WDM carriers.  

The receiver can use a linear adaptive equalizer to mitigate the performance penalty induced by the time-varying ISI. The effectiveness of the equalization will depend on how fast the dynamics of ${\bm{H}}_{n}^{(k)}$ can be tracked over time. Performance gains from inter-channel NLC have been observed in receivers using recursive least squares (RLS) \cite{Dar2015} equalizers and Kalman filters combined with maximum likelihood sequence estimation (MLSE) \cite{Golani2018_1}.

For the analysis presented in this work, the RLS algorithm is implemented by the complex-valued 2$\times$2 adaptive filter described in Eq. (\ref{RLS_filter}), whereas the update of the coefficients is performed using equations (\ref{Update_S}) and (\ref{Update_h}) \cite{Diniz2002}:

\begin{equation}\label{RLS_filter}
 \begin{bmatrix}\hat{A}_{x}(k) \\ \hat{A}_{y}(k) \end{bmatrix} = \begin{bmatrix} \bm{h}^H_{xx} & \bm{h}^H_{xy} \\ \bm{h}^H_{yx} & \bm{h}^H_{yy}\end{bmatrix}\!\! \begin{bmatrix} \bm{a}_{x}(k) \\ \bm{a}_{y}(k)\end{bmatrix}
\end{equation}

\begin{flalign} \label{Update_S}
\bm{S}_x(k+1) &= \frac{1}{\lambda}\left[\bm{S}_x(k) - \frac{\bm{S}_x(k)\bm{a}_{x}(k)\bm{a}_{x}(k)^H\bm{S}_x(k)}{\lambda + \bm{a}_{x}(k)^H\bm{S}_x(k)\bm{a}_{x}(k)} \right] \nonumber\\
\bm{S}_y(k+1) &= \frac{1}{\lambda}\left[\bm{S}_y(k) - \frac{\bm{S}_y(k)\bm{a}_{y}(k)\bm{a}_{y}(k)^H\bm{S}_y(k)}{\lambda + \bm{a}_{y}(k)^H\bm{S}_y(k)\bm{a}_{y}(k)} \right] 
\end{flalign}

\begin{flalign} \label{Update_h}
\bm{h}_{xx}(k+1) &= \bm{h}_{xx}(k) + e_{x}^*(k)\bm{S}_{x}(k+1)\bm{a}_{x}(k)\nonumber\\
\bm{h}_{xy}(k+1) &= \bm{h}_{xy}(k) + e_{x}^*(k)\bm{S}_{y}(k+1)\bm{a}_{y}(k)\nonumber\\
\bm{h}_{yx}(k+1) &= \bm{h}_{yx}(k) + e_{y}^*(k)\bm{S}_{x}(k+1)\bm{a}_{x}(k)\nonumber\\
\bm{h}_{yy}(k+1) &= \bm{h}_{yy}(k) + e_{y}^*(k)\bm{S}_{y}(k+1)\bm{a}_{y}(k)
\end{flalign}

\noindent where $\mathrm{N}$ is the number of filter taps, $\bm{a}_{x}(k)~=~[\hat{A}_x(k-d),...,\hat{A}_x(k-d-N)]^T$, $\bm{a}_{y}(k)~=~[\hat{A}_y(k-d),...,\hat{A}_y(k-d-N)]^T$ and $d$ is the decision delay. The filter components of the equalizer have the form $\bm{h}(k)~=~[h_0, h_1,..., h_{N-1}]^T$. $\bm{S}_{x}(k)$ and $\bm{S}_{y}(k)$ are $\mathrm{N\times N}$ matrices corresponding to the inverse of the deterministic correlation matrix of the symbols in each polarization, and $\lambda$ is the forgetting factor. Finally, $\bm{e}(k)~=~[e_{x}(k), e_{y}(k)]^T$ is the error between the outputs of the filter and the desired symbols.
\subsection{Proposed NLC Scheme}

\begin{figure}[t!]
\centering
\includegraphics[width=0.95\linewidth]{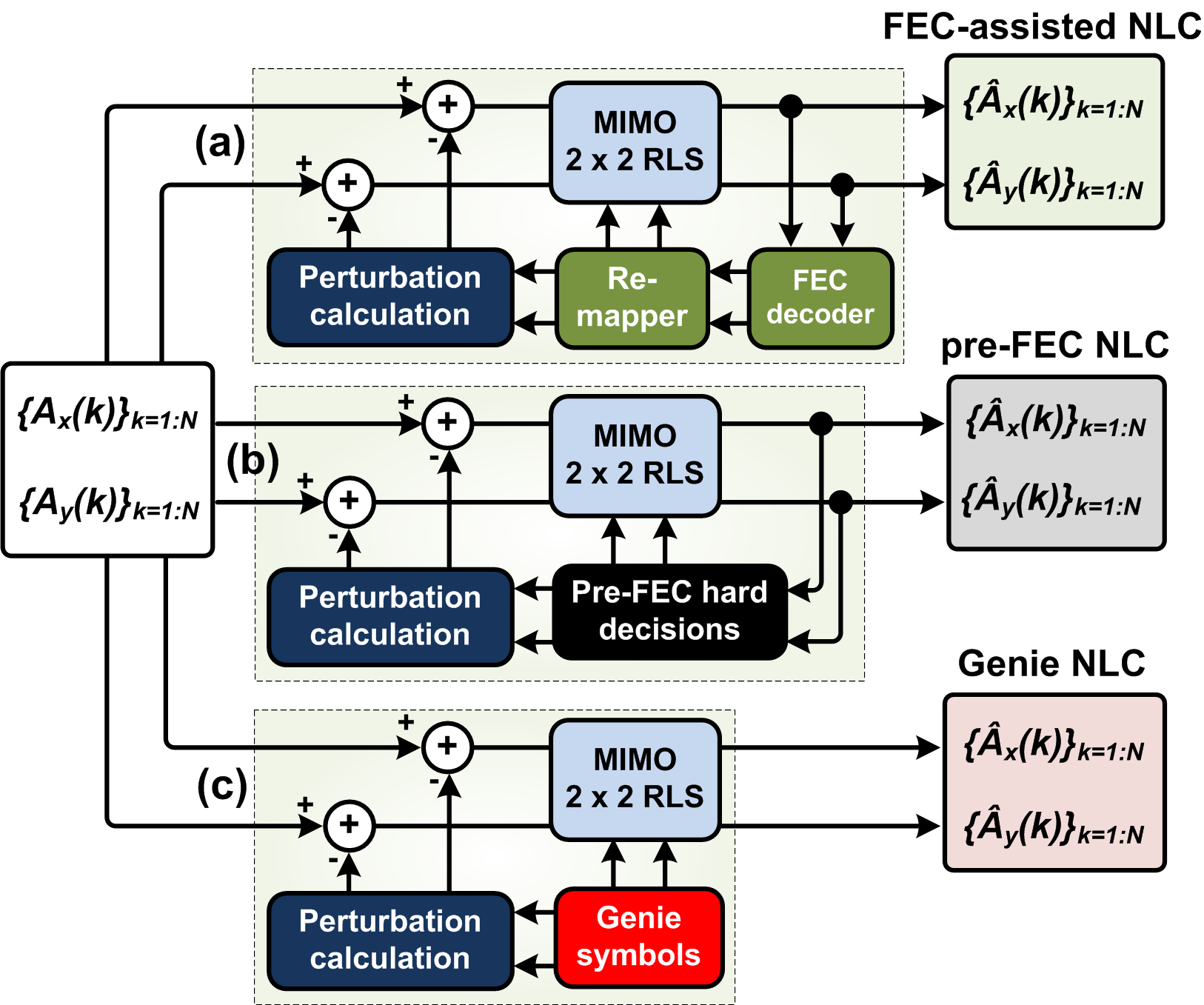}
\caption{NLC schemes investigated. (a) FEC-assisted; (b) Conventional; (c) Genie-assisted.}
\label{Fig1}
\end{figure}

The performance of DFE equalizers suffers degradation due to propagation of errors in the decision stage. Since the FEC in coherent WDM systems is designed to allow reliable communication even when the pre-FEC BERs are as high as $10^{-2}$, the perturbation-based intra-channel NLC will suffer performance degradation when the receiver operates in the range of SNR close to the pre-FEC BER limits. Similar comments can be made about the performance of RLS filters used to compensate inter-channel NLI. Alternatively, to mitigate this problem we assume that the receiver may use a feedback from the FEC decoder attempting to improve the NLI estimation (Fig.\ref{Fig1}(a)) and, thereby, the NLC performance.

In the proposed iterative method, at each iteration, an updated estimate of the intra-channel NLI is calculated based on a sequence of symbols regenerated from the output of the FEC decoder (in this work - LDPC). For comparison, we also evaluate the performance of an idealized genie-assisted perturbation NLC, where all the transmitted symbols are known \textit{a priori} at the receiver. Additionally, we also investigate the receiver performance when an RLS linear adaptive equalizer is included within the iterative processing, with the task of compensating for the fractions of the time-varying inter-channel and residual intra-channel NLI that are slow enough to be tracked (see Section \ref{InterNLC}). In the FEC-assisted mode, the error used to update RLS coefficients is calculated with respect to the output of the decoder.

In this paper, we focus on comparing the schemes in Fig.\ref{Fig1}(a)-(b) with the ideal genie-assisted scheme shown in Fig.\ref{Fig1}(c), where the NLI is compensated assuming full knowledge of the transmitted symbols.
 
 \begin{figure*}[t!]
\centering
\includegraphics[width=1.0\linewidth]{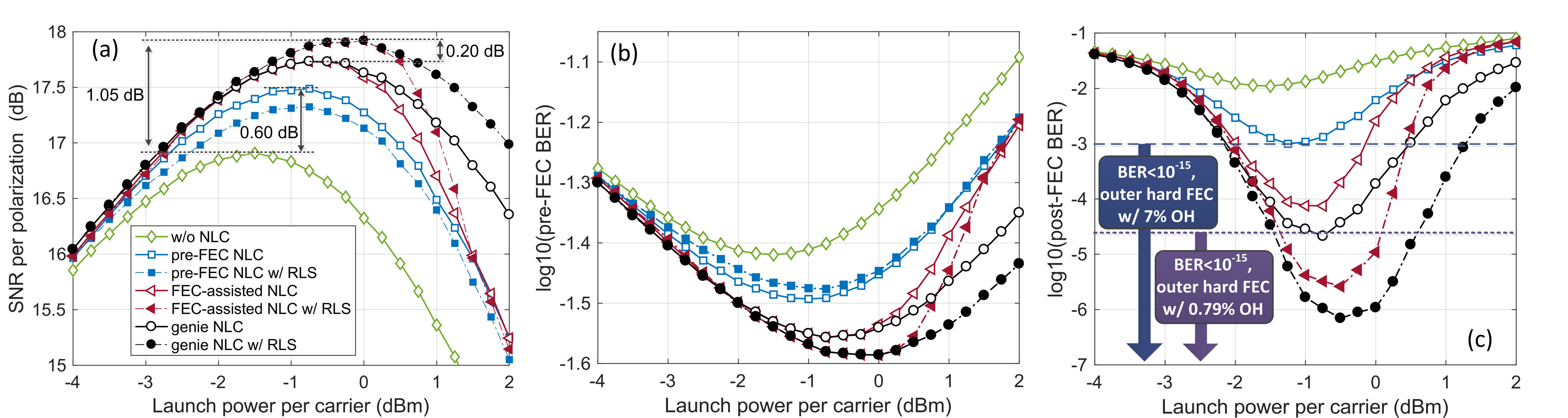}
\caption{Numerical results for the receiver performance as a function of launched power after 20$\times$80 km of dispersion uncompensated WDM transmission of 5 $\times$ 32~GBd DP-64QAM. (a) SNR at the input of the LDPC decoder; (b) pre-FEC BER; (c) post-FEC BER.}
\label{Fig2}
\end{figure*}

\section{Numerical Simulations}\label{SecIII}
The receivers detailed in Fig.\ref{Fig1} are firstly analyzed in Monte Carlo simulations with the SSFM. The simulation model considers a WDM system composed of five carriers modulated at 32~GBd and separated in a grid spacing of 37.5~GHz. The transmitted bits are generated by encoding a pseudo random bit sequences with a LDPC code of rate $\mathrm{R=5/6}$ (20\% overhead DVB-S.2 standardized FEC). The encoded bits are interleaved and Gray mapped to a DP-64QAM symbols. For each Monte Carlo run, each polarization signal carries four LDPC blocks of 64800 encoded bits per WDM carrier. The signal of each carrier is upsampled to 16~samples/symbol and pulse shaped with a root-raised cosine (RRC) filter with 401~taps and roll-off factor of 0.005. 

The transmission link model corresponds to 20$\times$80~km spans of SMF, with all losses compensated by Erbium-doped fiber amplifiers (EDFAs) with noise figure of 4.5~dB. The nonlinear propagation is simulated with the SSFM at a fixed step-size of 100~m (800~steps/span). The fiber parameters attenuation, nonlinear coefficient and chromatic dispersion are set to be $\mathrm{\alpha} =$~0.2~$\mathrm{dB/km}$, $\mathrm{\gamma} =$~1.~3~$\mathrm{W^{-1}km^{-1}}$, and $\mathrm{D} =$~17~$\mathrm{ps/nm/km}$, respectively. Polarization effects, such as PMD, were not included in the simulations. 

At the receiver, the signal passes through CD compensation, low-pass filtering, decimation to 2~samples/symbol, $\mathrm{T_s/2}$-fractionally spaced minimum mean square error (MMSE) equalization (24~taps). The estimated symbols are then sent to the iterative stage where the first order perturbation model and the RLS filter are used to perform intra- and inter-channel NLC, respectively. The matrix of coefficients $\mathrm{C}$ is calculated assuming a fixed memory length $\mathrm{L}$ of 80 symbols. In order to reduce the complexity of the data processing, a cutoff threshold of -~16~dB is chosen to discard coefficients much smaller than $\mathrm{C(0,0)}$. The choice of $\mathrm{L}$ was based on a coarse optimization of the NLC performance, whose saturation point was observed for $\mathrm{L}\approx$ 80 symbols. The RLS adaptive equalizer is configured with 5~taps and forgetting factor ranging within the interval [0.98,1]. The LDPC decoder is configured to perform a fixed number of 10 decoding iterations per block.

For all results shown in this paper, SNR always refers to effective received SNR, which is calculated using the following estimator 

\begin{equation}\label{Eq5}
\mathrm{SNR} \approx \frac{1}{N}\sum_{k=1}^{N}\frac{|A(k)|^2}{|\hat{A}(k)-A(k)|^2},
\end{equation}
\noindent where $A(k)$ and $\hat{A}(k)$ is a pair of the transmitted and the corresponding received symbol of a training sequence of length $\mathrm{N}$.

Figure~\ref{Fig2} shows the results obtained after extensive numerical simulations. For each launch power, the BER values shown correspond to the average BER over all WDM carriers and over at least ten Monte Carlo runs, corresponding to at least $\mathrm{2.13\times 10^7}$ information bits in total. The performance without NLC is included for comparison. 

The average SNR per polarization is shown in Fig.~\ref{Fig2}~(a) for the three receivers depicted Fig.~\ref{Fig1}. Without the RLS filter, for the pre-FEC NLC scheme, the maximum SNR is increased by 0.60~dB, whereas for the FEC-assisted NLC scheme an additional gain of $\approx$~0.25~dB is obtained. Adding the RLS filter, the FEC-assisted scheme exhibits a further improvement of 0.2~dB, whereas the gain of the pre-FEC NLC scheme is penalized by $\approx$~0.15~dB. This penalty is due to the fact that the RLS is using pre-FEC hard decisions to quickly adapt the filter taps and, therefore, the increased number of wrong symbol decisions influences the ability of the equalizer to tracking fast time-varying ISI generated from the inter-channel NLI, as compared to the FEC-assisted NLC scheme. It was found that this penalty vanishes by choosing higher values for the forgetting factor. More interestingly, the performance of the FEC-assisted NLC scheme is similar to the performance of the genie-assisted NLC scheme for a number of points, including the optimal launch power.

The translation of SNR into pre-FEC and post-FEC BER is shown in Fig.~\ref{Fig2}~(b)-(c), respectively. It is noted that, even though pre-FEC BER follow a similar pattern observed in the SNR, the post-FEC performance of the FEC-assisted NLC scheme deviates from the genie-assisted NLC curve. A possible reason for this behavior can be related to difference on decision error distributions of the symbols produced after FEC-assisted NLC and genie-assisted NLC. 

It is observed that around the optimal launch power, the receiver requires a maximum of three iterations between FEC decoder and equalization to achieve the minimum BER. Overall, a gain of $\approx$~1.0~dB of SNR per polarization is obtained by the proposed scheme with respect to the performance without NLC, reducing the BER after the LDPC decoder from $\mathrm{10^{-2}}$ to less than $\mathrm{10^{-5}}$.

In the next section, the analysis is extended to investigate if the performance characterization obtained by numerical simulations can be verified in transmission experiments.

\section{Experimental Analysis}\label{SecIV}

\begin{figure}[b!]
\centering
\includegraphics[width=1.0\linewidth]{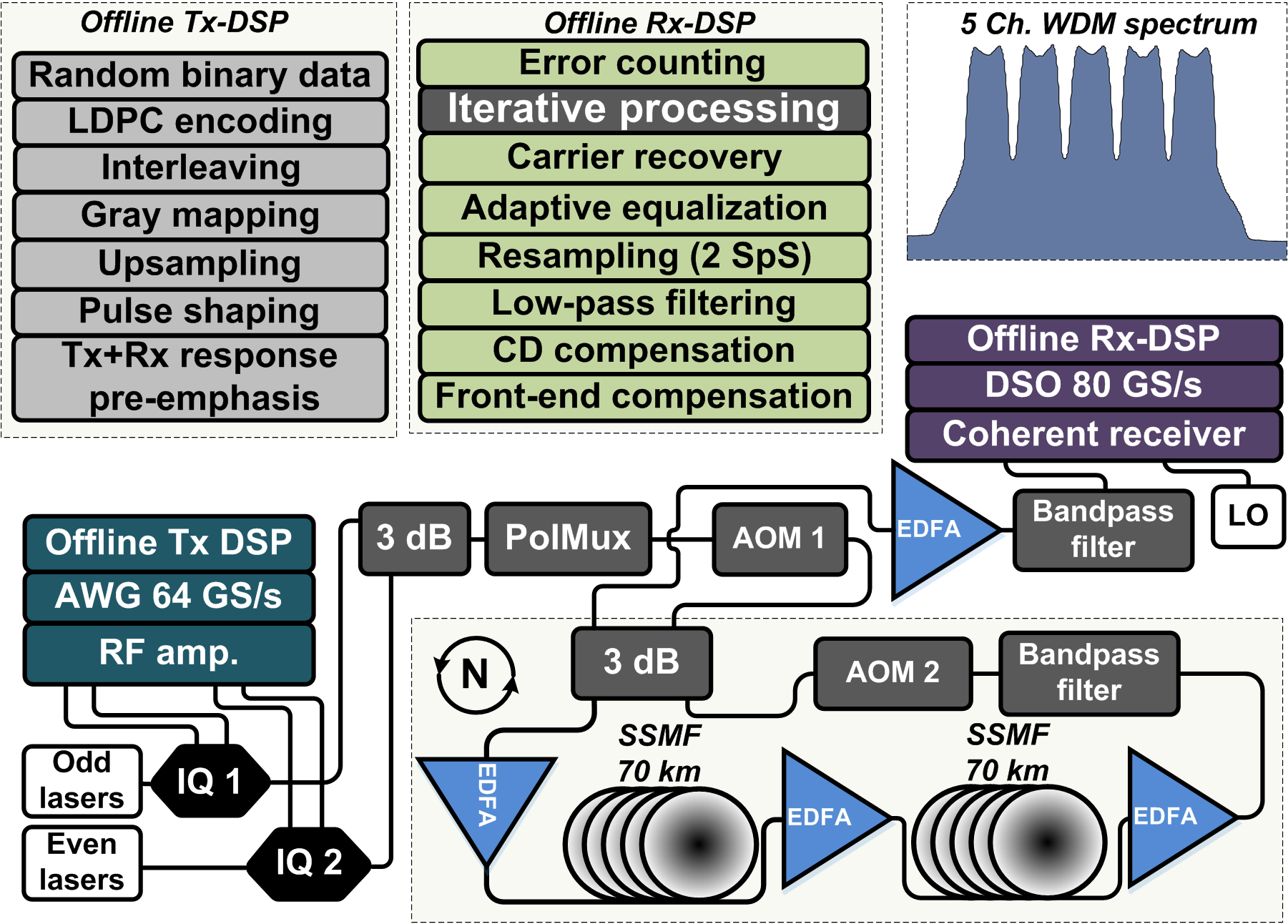}
\caption{Experimental setup with the detailed digital signal processing at the transmitter and at the receiver.}
\label{Fig3}
\end{figure}

\begin{figure*}[t!]
\centering
\includegraphics[width=1.0\linewidth]{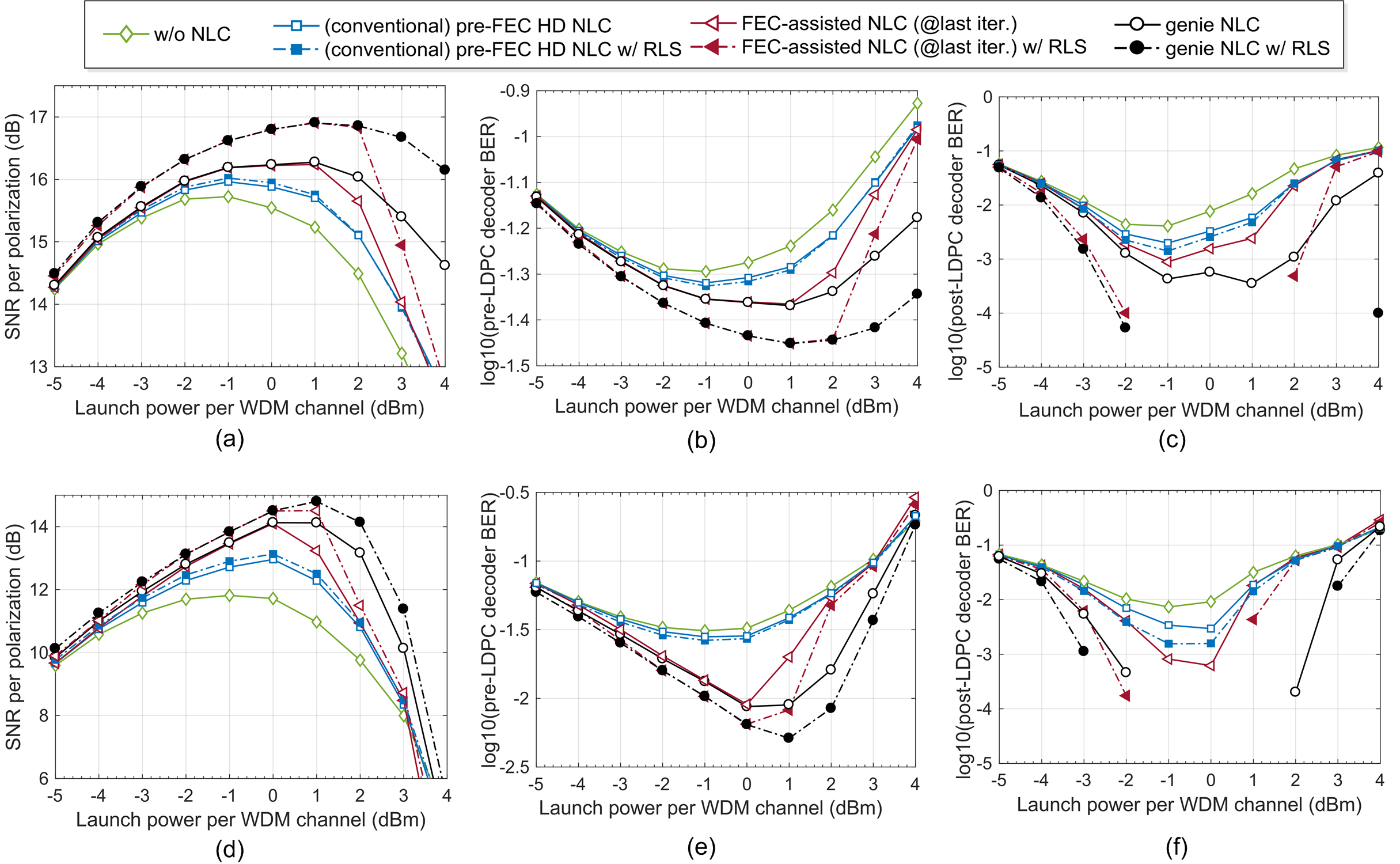}

\caption{Experimental results for the central channel performance as a function of launched power for dispersion uncompensated WDM transmission of (a-c)  5 $\times$ 32~GBd DP-64QAM after 1120~km and (d-f) 5 $\times$ 32~GBd DP-16QAM after 4200~km.}
\label{Fig4}
\end{figure*}

The experimental setup is shown in Fig.~\ref{Fig3}. The WDM system is composed of five carriers modulated at 32~GBd and disposed in a grid spacing of 50~GHz. The transmitted symbols are generated by encoding an pseudo random bit sequences with LDPC code rates $\mathrm{R=5/6}$ (20\% overhead) for DP-16QAM and $\mathrm{R=3/4}$ (33\% overhead) for DP-64QAM (DVB-S.2 standardized FEC). The encoded bits are interleaved and Gray mapped to QAM symbols. Two decorrelated sequences of eight LDPC blocks (64800 encoded bits per block) are loaded in the arbitrary waveform generator (AWG). The signal is pulse shaped with a root-raised cosine (RRC) filter with 401~taps and roll-off factor of 0.5. A linear pre-emphasis is applied in order to compensate for the combined frequency response of transmitter and receiver. After amplification, each baseband signal drives one of two in-phase/quadrature (IQ) modulators. The even-odd five carrier WDM system is obtained after further combination in a polarization multiplexing stage. All optical carriers in the experiment are external cavity lasers with 10~kHz linewidth. 

In back-to-back configuration, the maximum effective received SNR of the central WDM channel saturates at 20.5~dB. The WDM channels propagate in a recirculating loop composed of two 70~km spans of standard single mode fiber (SSMF), with all the losses compensated by EDFAs. After coherent detection, the signal passes through a front-end compensation stage, resampling, CD compensation, low-pass filtering, decimation, $\mathrm{T_s/2}$-fractionally spaced adaptive equalization (85~taps, trained blindly and with 5\% pilot-symbols for 16QAM and 64QAM, respectively), and carrier recovery with a digital direct-decision phase-locked loop. 

The estimated symbols are sent to the iterative stage where intra- and inter-channel NLC are performed. As in Section~\ref{SecIII}, the matrix of coefficients $\mathrm{C}$ is calculated assuming fixed memory length $\mathrm{L}$ of 80 symbols and a cutoff threshold of -16~dB to discard coefficients much smaller than $\mathrm{C(0,0)}$. Note that larger $\mathrm{L}$ values could be required to maximize the NLC performance as the transmission distance increases. However, in the processing of the experimental data it was noted that, with $\mathrm{L}=$ 80, most of the gain observed in the numerical simulations was achieved for the transmission distances of interest. Hence, for simplicity, the parameter $\mathrm{L}$ was chosen the same as for the processing of the numerical simulations.  The RLS adaptive equalizer is configured with 3~taps and forgetting factor varying within the interval [0.98,1]. For each time decoding is attempted, the LDPC decoder performs a fixed number of 5 and 10 decoding iterations per block of symbols when processing 16QAM and the 64QAM, respectively.

\subsection{Performance of the Central Channel}\label{SecIVsubA}

The experimental results are shown in Figure~\ref{Fig4}. For each launch power, the BER values shown correspond to the average BER of the central WDM channel over 96 FEC blocks. The performance without NLC is included for comparison. Figures~\ref{Fig4}~(a, d) show the average effective received SNR per polarization for the three receivers depicted in Fig.\ref{Fig1}. Thereafter, the SNR results for DP-64QAM are followed by the results for DP-16QAM in parenthesis. 

First the performance is evaluated only for intra-channel NLC, i.e. without the RLS filter in the NLC scheme. In this case, the maximum received SNR is increased by 0.20~dB (1.0~dB) with the pre-FEC NLC scheme, whereas using the FEC-assisted iterative scheme an extra gain of $\approx$~0.20~dB (1.0~dB) is obtained. Including the RLS filter, the FEC-assisted scheme provides an additional improvement of 0.6~dB (0.5~dB), whereas the gain of the pre-FEC NLC scheme remains the same. It is clear that, in both cases, the performance of the RLS filter is enhanced by the FEC-assisted scheme. Similarly to the simulation results, the performance of the FEC-assisted NLC scheme is close to the performance of the genie-assisted NLC scheme for a number of points, including the optimal launch power. 

Figures~\ref{Fig4}~(b,c)-(e,f) show the pre-FEC and post-FEC BER performance, respectively. It is seen that, even though pre-FEC BER follow a similar pattern observed in the SNR, the post-FEC performance of the FEC-assisted NLC scheme deviates from the genie-assisted NLC curve, as also noted in the simulation results. Around the optimal launch power, the receiver required around three iterations between FEC decoder and NLC to achieve the minimum BER, whereas in the nonlinear regime the number of iterations increases. All the results displayed here correspond to a fixed number of five iterations between LDPC decoder and the NLC. 

An aggregated increase of $\approx$~1.0~dB (2.5~dB) of received SNR per polarization is obtained by the proposed scheme with respect to the performance without NLC, lowering the post-LDPC decoder BER from $\mathrm{10^{-2}/10^{-3}}$ to less than $\mathrm{5\times 10^{-5}}$. Assuming that an outer linear hard FEC code is used to bring the BER down to below $\mathrm{10^{-15}}$, error-free performance can be achieved with an extra overhead of 0.79\%, i.e. assuming a pre-hard-FEC limit BER of $\mathrm{5\times 10^{-5}}$ \cite{Millar2016}. 

\subsection{WDM Performance and Comparison with Single-Channel DBP}\label{SecIVsubB}

\begin{figure*}[t!]
\centering
\includegraphics[width=1.0\linewidth]{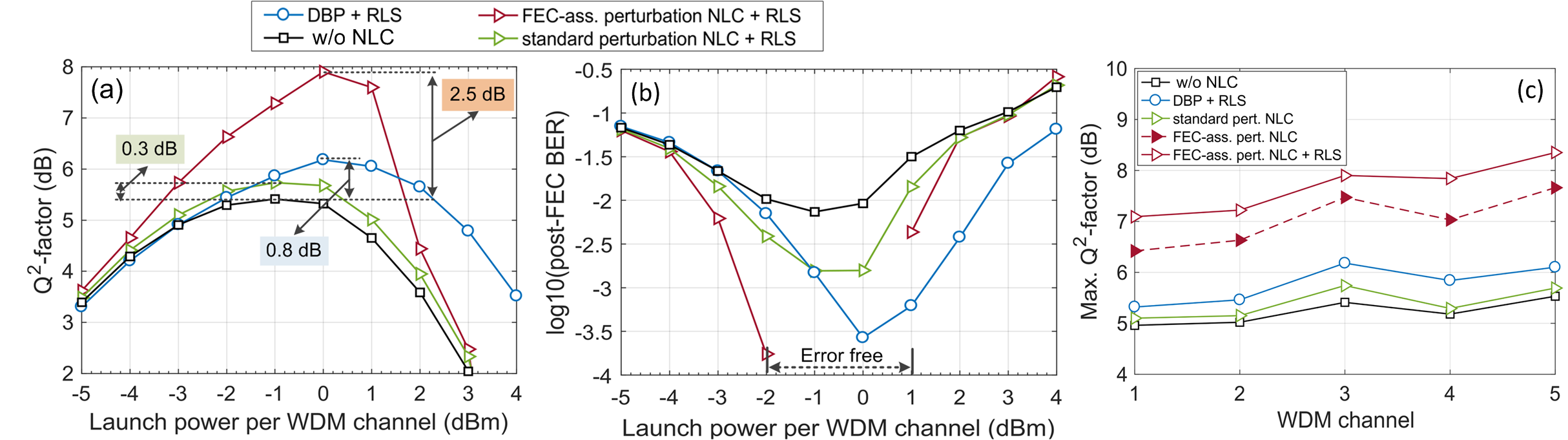}
\caption{Experimental performance comparison of DBP vs Perturbation-based FEC-assisted NLC. (a) pre-FEC $\mathrm{Q^2}$-factor of the central channel as a function of the launch power; (b) post-FEC BER of the central channel as a function of the launch power; (c) Maximum pre-FEC $\mathrm{Q^2}$-factor ($\mathrm{Q^2}$-factor at the optimum launch power) for each WDM carrier.}
\label{Fig5}
\end{figure*}
Here we focus on the long-haul WDM transmission of 5$\times$32~GBd DP-16QAM to compare the performance of the proposed perturbation-based NLC schemes with the conventional single-channel DBP. The DBP algorithm is implemented with a symmetric SSF method assuming the Manakov model for signal propagation \cite{Ip2010}. The algorithm runs with a constant step-size of 1~km and a sampling rate of 2~samples/symbol (64 GS/s). The choice of step size is done to guarantee that the DBP algorithm will operate at the best performance for the case under study. The attenuation, chromatic dispersion and nonlinear coefficients assumed by the algorithm are fine tuned to maximize the performance of the NLC at the optimum launch power. 

In Fig.~\ref{Fig5}(a) the pre-FEC $\mathrm{Q^2}$-factor of the central channel as a function of the power launched into the fiber is shown for different DSP configurations at the receiver. Here the pre-FEC $\mathrm{Q^2}$-factor is shown because it is the most popular figure of merit to evaluate performance gains obtained by DBP. A $\mathrm{Q^2}$-factor gain of 0.3~dB is obtained by applying the standard perturbation NLC, whereas DBP is able to provide a gain of 0.8~dB. The gain observed for the FEC-assisted perturbation NLC is 2.5~dB. 

In Fig.~\ref{Fig5}(b) the post-FEC BER as a function of the launch power is displayed. As discussed in Section~\ref{SecIVsubA}, under the assumption that the receiver uses an outer hard FEC with a small overhead to bring the BER down to below $\mathrm{10^{-15}}$, \quotes{error free} performance is achieved only by the iterative FEC-assisted perturbation NLC. All the non-zero BER values correspond to the performance of the system after a maximum of 5 iterations (stopping criterion) between NLC and decoder, whereas both \quotes{error free} points were obtained after 2 iterations.

It is interesting to note that for 1~dBm of launch power per channel, in the highly nonlinear regime, the post-FEC BER of DBP is the lowest, despite the fact that the pre-FEC $\mathrm{Q^2}$-factor of the FEC-assisted perturbation NLC is more than 1.0~dB higher. The origin of this results is currently under investigation, however it is probably  related to the fact that DFE equalizers may generate bursts of symbol errors that could deteriorate the performance of the LDPC decoder. As also highlighted in \cite{Silva2018}, the performance of the FEC-assisted perturbation NLC approaches the performance of the standard perturbation NLC method in the highly nonlinear regime. 

The maximum pre-FEC $\mathrm{Q^2}$-factor for all WDM channels is shown in Fig.~\ref{Fig5}(c). The difference in performance between channels is mostly due to the tilt of the amplification noise power density accumulated over 4200~km (30 loop turns). Nevertheless, the performance gain of each NLC scheme is approximately uniform for all measured channels. The FEC-assisted perturbation NLC outperforms the standard perturbation NLC and DBP for all cases.

\section{Conclusion}
The performance of a perturbation-based intra and inter-channel nonlinearity compensation (NLC) scheme was investigated via numerical simulations and transmission experiments. The proposed scheme enhances the performance of the receiver by using iterations between NLC algorithms and an LDPC decoder. Experimental results show that FEC-assisted NLC outperforms pre-FEC NLC, improving the bit error rate performance of a 5$\times$32~GBd WDM system with DP-16QAM and DP-64QAM and after 4200~km and 1120~km, respectively, of dispersion uncompensated transmission.

\section{Acknowledgements}

This work was supported by the Danish National Research Foundation (DNRF) Research Centre of Excellence, SPOC, ref. DNRF123.

\bibliographystyle{abbrv}

\end{document}